\documentclass[aps,prb,superscriptaddress,reprint,showpacs,twocolumn,preprintnumbers,floatfix]{revtex4-2}
\usepackage[dvips]{graphicx}
\usepackage[latin1]{inputenc}
\usepackage{xcolor,bm}
\usepackage{multirow}
\usepackage{hyperref}
\hypersetup{
	colorlinks=true,
	linkcolor=blue,
	filecolor=blue,
	citecolor=blue,      
	urlcolor=blue,
}
\usepackage{multirow,easyReview}
\usepackage{amsmath}
\usepackage[charter]{mathdesign}
\usepackage[normalem]{ulem}

\begin{document}
\draft

\hyphenation{a-long}

\title{Cluster charge-density-wave glass in hydrogen-intercalated TiSe$_{2}$}

\author{Giacomo~Prando}\email{giacomo.prando@unipv.it}\affiliation{Dipartimento di Fisica, Universit\`a degli Studi di Pavia, I-27100 Pavia, Italia}
\author{Erik~Piatti}\affiliation{Department of Applied Science and Technology, Politecnico di Torino, I-10129 Torino, Italia}
\author{Dario~Daghero}\affiliation{Department of Applied Science and Technology, Politecnico di Torino, I-10129 Torino, Italia}
\author{Renato~S.~Gonnelli}\affiliation{Department of Applied Science and Technology, Politecnico di Torino, I-10129 Torino, Italia}
\author{Pietro~Carretta}\affiliation{Dipartimento di Fisica, Universit\`a degli Studi di Pavia, I-27100 Pavia, Italia}

\widetext

\begin{abstract}
	The topotactic intercalation of transition-metal dichalcogenides with atomic or molecular ions acts as an efficient knob to tune the electronic ground state of the host compound. A representative material in this sense is 1$T$-TiSe$_{2}$, where the electric-field-controlled intercalations of lithium or hydrogen trigger superconductivity coexisting with the charge-density wave phase. Here, we use the nuclear magnetic moments of the intercalants in hydrogen-intercalated 1$T$-TiSe$_{2}$ as local probes for nuclear magnetic resonance experiments. We argue that fluctuating mesoscopic-sized domains nucleate already at temperatures higher than the bulk critical temperature to the charge-density wave phase and display cluster-glass-like dynamics in the MHz range tracked by the {}$^{1}$H nuclear moments. Additionally, we observe a well-defined independent dynamical process at lower temperatures that we associate with the intrinsic properties of the charge-density wave state. In particular, we ascribe the low-temperature phenomenology to the collective phason-like motion of the charge-density wave being hindered by structural defects and chemical impurities and resulting in a localized oscillating motion.
\end{abstract}

\date{\today}

\maketitle

\narrowtext

\section{Introduction}\label{SectIntro}

The wide range of electronic properties characterizing the transition-metal dichalcogenides (TMDs) makes these materials remarkably interesting from both fundamental and application-oriented perspectives \cite{Rad11,Wan12,Gei13,Kop14,Man17}. The representative chemical formula of TMDs is MX$_{2}$, M and X being transition-metal and chalcogen elements, respectively. In spite of their vast chemical variety, TMDs share a common crystallographic structure resulting from the stacking of MX$_{2}$ trilayers. Each trilayer is composed of an atomically-thin layer of M ions sandwiched between two closely-packed layers made of X ions covalently-bonded with the M ions \cite{Wil69}. The stacked trilayers interact among them via weak van-der-Waals-like forces, the covalent-bond-free interface between different trilayers being referred to as the van der Waals gap. As a result, TMDs can be conveniently exfoliated down to the single-trilayer, two-dimensional limit. The layered structure makes TMDs suited hosts for topotactic intercalation with atomic or molecular ions as well as neutral molecules \cite{Dre86}. The interest for intercalated TMDs is their potential as solid ionic conductors \cite{Whi78,Sam89,Pow93} and, at the same time, the efficient control on the electronic properties of the host material offered by the intercalation \cite{Par79,Fri87,Kle15,Voi15,Wan18,Li21,Wan21,Van22}.

A prototypal example in this sense is 1$T$-TiSe$_{2}$. The pristine composition has been investigated extensively since the 1970's after the first reports of a commensurate $2 \times 2 \times 2$ charge-density wave (CDW) state developing from a high-temperature semimetallic phase \cite{DiS76,Ras08}. The historical relevance of this discovery, together with similar observations in other TMDs, was the departure from the then-accepted paradigm, due to Peierls, of the CDW transition being specific of one-dimensional systems simultaneously undergoing a metal-to-insulator phase transition -- both aspects being violated in TMDs \cite{Pei55,DiS79,Col88,Gru88,Tho96,Wil01,Li07,Mon12,Zhu17}. Although the transition temperature to the CDW phase is $T_{\textrm{CDW}} \sim 200$ K, a gap opening was reported in limited portions of the Fermi surface already at higher temperatures \cite{Miy95,Che16}, consistently with the observation of higher transition temperatures to the CDW state in the single-trilayer limit \cite{Gol12,Che15,Sug16}. The properties of the CDW phase in pristine 1$T$-TiSe$_{2}$ are highly peculiar \cite{Ish10,Xu20} and, almost fifty years after its first observation, the discussion on the main driving microscopic mechanism underlying such state -- electron-phonon coupling as opposed to exciton condensation -- is still ongoing \cite{Hal68a,Hal68b,Hug77,Whi77,Wil77,Gab81,Hol01,Kid02,Ros02,Kog17,Weg20,Ott21,Lin22,Nov22}.

The electronic properties of the pristine 1$T$-TiSe$_{2}$ composition can be tuned by physical and/or chemical means. Superconductivity with critical temperatures around a few Kelvins is triggered by pressure \cite{Kus09}, electric-field-induced charge doping \cite{Li16}, and by intercalation using different atomic ions \cite{Mor06,Lia21,Pia22}. The CDW phase is suppressed by electric-field-induced charge doping \cite{Li16}, copper intercalation \cite{Mor06}, and titanium self-doping \cite{Jao19} -- however, the electric-field-controlled intercalation using lithium \cite{Lia21} and hydrogen \cite{Pia22} does not alter the $T_{\textrm{CDW}}$ value estimated by means of electrical transport measurements. Remarkably, a structure of nanoscopic domains with alternating commensurate-incommensurate CDW phases emerges in 1$T$-TiSe$_{2}$ when perturbed by electric fields or by the Li intercalation, the superconducting state being possibly confined to the incommensurate-CDW domains \cite{Li16,Lia21}. In spite of the detailed description of the spatial intertwining of different phases at low temperatures, a microscopic investigation of the high-temperature CDW state in the intercalated systems is still missing.

In this work, we use the magnetic moments of {}$^{1}$H nuclei in hydrogen-intercalated 1$T$-TiSe$_{2}$ as local probes for nuclear magnetic resonance (NMR) experiments. Based on our modelling of the temperature dependence of the spin-lattice relaxation rate of the {}$^{1}$H nuclear magnetization, we argue that a cluster CDW glass state is realized in the high-temperature limit, favoured by the disordered configuration of the intercalants. Our evidences suggest that fluctuating mesoscopic-sized domains nucleate already at temperatures higher than $T_{\textrm{CDW}}$, slowing down towards a percolating CDW state throughout the sample and resulting in slow dynamics in the MHz range tracked by the {}$^{1}$H nuclear moments. Additionally, we observe a well-defined independent dynamical process at lower temperatures that we associate with the intrinsic properties of the CDW state. In particular, we argue that the collective phason-like motion of the CDW is hindered by structural defects and chemical impurities acting as pinning centres, resulting in a localized oscillating motion detected by the nuclear moments.

\section{Experimental details and results}\label{SectExp}

We protonated flake-like 1$T$-TiSe$_{2}$ single crystals (HQ Graphene) immersing them in a Duran crucible filled with $1$-ethyl-$3$-methylimidazolium tetrafluoroborate (Sigma Aldrich) immediately after cleavage and applying a gate voltage $3$ V between a platinum electrode and 1$T$-TiSe$_{2}$ \cite{Pia22}. For this aim, the crystals were electrically contacted via a small droplet of silver paste (RS Components). All the samples were gated in ambient atmosphere and at temperature $T = 300$ K, using the gating time as a knob to control the protonation state \cite{Pia22}. After the gating process, we thoroughly rinsed the crystals with acetone and ethanol and stored them in a vacuum dessicator.

We performed pulsed {}$^{1}$H NMR experiments \cite{Abr61,Fuk81,Ern87,Sli90,Lev08} at two fixed values of magnetic field ($\mu_{0}H = 1.95$ T and $3.46$ T). The field was oriented parallel to the crystallographic $ab$ planes of a collection of six stacked protonated 1$T$-TiSe$_{2}$ crystals (gating time $60$ minutes), each with approximate surface $\sim 2 \times 2$ mm$^{2}$. We controlled the sample temperature with a dynamic continuous-flow cryostat using either liquid helium or liquid nitrogen, quantifying the temperature value using a thermocouple end close to the sample site. The data shown in the text belong to several cooling and warming cycles and, in order to maximize the reproducibility of the results, we always aimed at comparable thermal protocols. In particular, we cooled the sample in-field down to the lowest attainable temperature (depending on the cryogenic liquid being used) and settled the temperature on warming to perform each measurement.

\begin{figure}[t!] 
	\centering
	\includegraphics[width=0.5\textwidth]{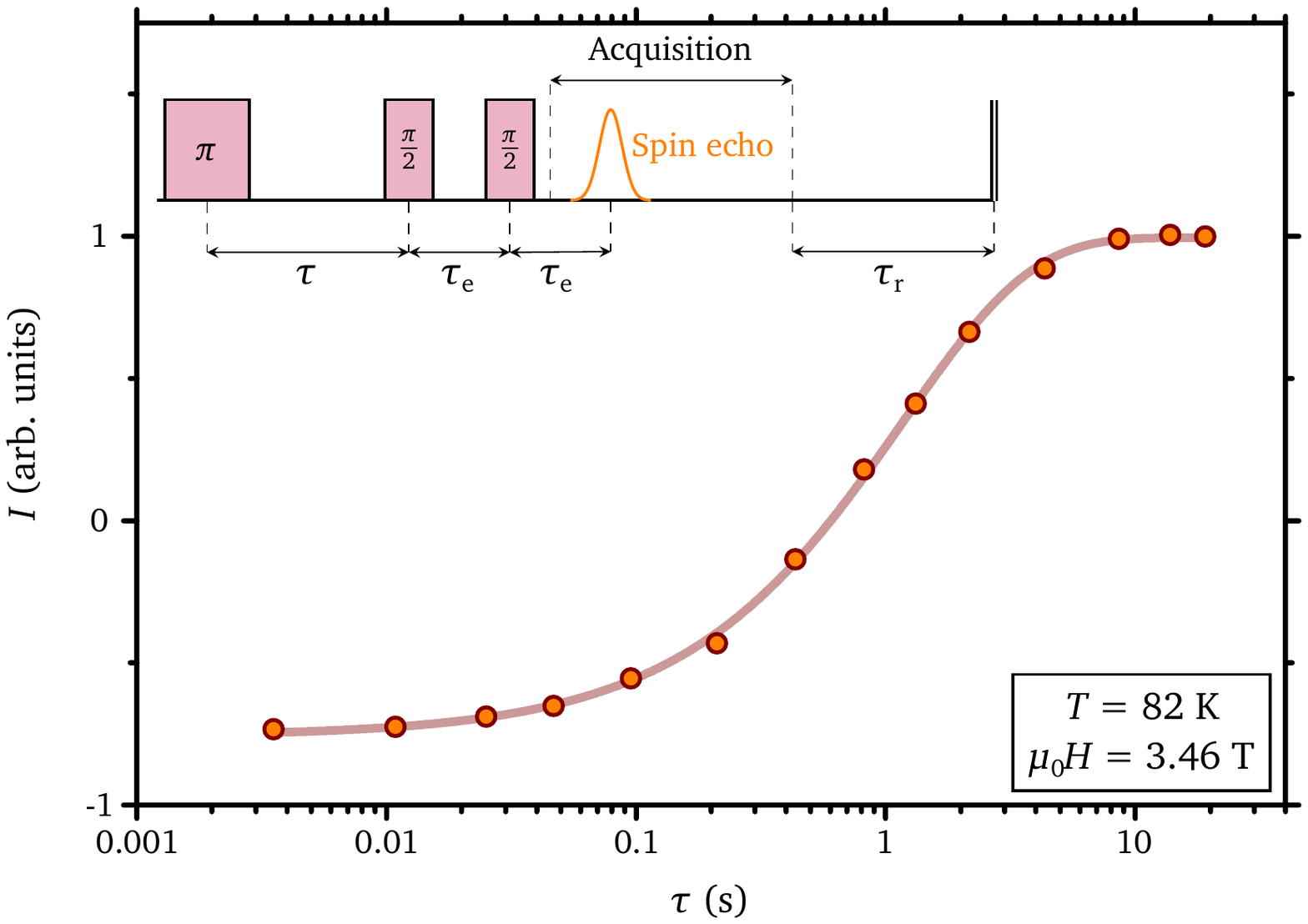}
	\caption{\label{RawRecData} Representative recovery of the {}$^{1}$H nuclear magnetization towards the thermodynamical equilibrium as a function of the time $\tau$ elapsed between the $\pi$ pulse and the first $\pi/2$ pulse. $I$ represents the integral of the fast-Fourier-transformed echo shown later in the inset to Fig.~\ref{FWHMvsT}. The continuous line is a best-fitting function according to Eq.~\eqref{EqRecoveryFitting}. The inset shows a scheme of the pulsed inversion sequence used to perform the measurements discussed throughout the text (time durations are not in proportion).}
\end{figure}
\begin{figure}[t!] 
	\centering
	\includegraphics[width=0.5\textwidth]{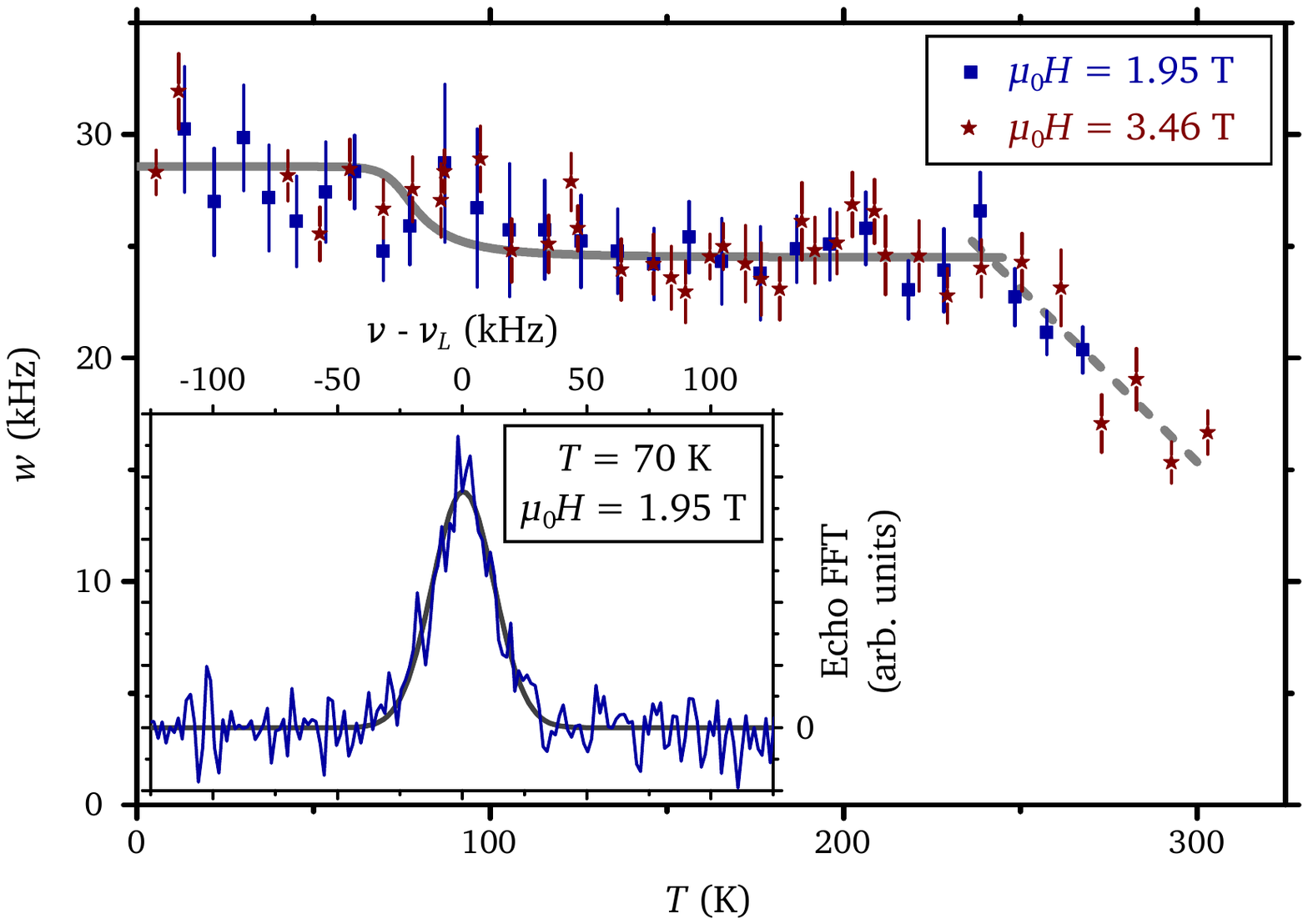}
	\caption{\label{FWHMvsT} Inset: the blue curve is the fast Fourier transform (FFT) of a representative echo signal at long recovery times. The black curve is a best-fitting of the data using the function in Eq.~\eqref{EqGaussian}. Main panel: dependence of the spectral width on temperature. The continuous line is obtained from Eq.~\eqref{EqAbragam} (see the discussion later on in the text). The dashed line is a guide to the eye.}
\end{figure}

We base all our results on the conventional inversion-recovery sequence of radiofrequency pulses shown in Fig.~\ref{RawRecData} (inset). The two $\pi/2$ pulses, with typical length $\sim 1 \; \mu$s and separated by $\tau_{e} \simeq 30 \; \mu$s, generate a solid echo at time $\tau_{e}$ after the second pulse. The sign and amplitude of the echo depend on the variable time $\tau$ between the inversion $\pi$ pulse and the first $\pi/2$ pulse. For each $\tau$ value, the sequence is repeated and the signal cumulated until a satisfactory signal-to-noise ratio is achieved. The second half of the echo signal is fast-Fourier-transformed and the integral $I$ of the transform is reported as a function of $\tau$ to obtain the time-recovery of the nuclear magnetization towards the state of thermal equilibrium -- see Fig.~\ref{RawRecData} for representative data. In the case of spin-$1/2$ nuclei, the expected recovery for the nuclear magnetization $M$ is purely exponential according to the following function \cite{Sli90}
\begin{equation}\label{EqRecoveryFitting}
	M\left(\tau\right) = M(0) + \Delta M \; \left\{1-\exp\left[-\left(\frac{\tau}{T_{1}}\right)^{\beta}\right]\right\}
\end{equation}
from which the spin-lattice relaxation time $T_{1}$ is quantified. The fitting function in Eq.~\eqref{EqRecoveryFitting} accounts for possible deviations of the pulse lengths from their ideal values, resulting in an imperfect initial inversion of the nuclear magnetization. Also, Eq.~\eqref{EqRecoveryFitting} accounts for possible deviations from the expected exponential behaviour through the empirical stretching parameter $\beta$, so accounting for non homogeneous local environments probed by the nuclei. Finally, the fast-Fourier-transformed echo for the longest $\tau$ value, i.e., such that the thermal equilibrium is attained, is fitted using an empirical Gaussian function
\begin{equation}\label{EqGaussian}
	y = \frac{A}{w} \sqrt{\frac{2}{\pi}} \exp\left[-2 \left(\frac{\nu-\nu_{L}}{w}\right)^{2}\right]
\end{equation}
in order to quantify the linewidth $w$ -- see the inset to Fig.~\ref{FWHMvsT} for representative data ($\nu_{L}$ is the Larmor frequency and $A$ the subtended area). Importantly, the repetition time $\tau_{r}$ is calibrated in every measurement so that $\tau_{r} \gtrsim 5 \; T_{1}$ to allow the nuclear magnetization to reach the equilibrium state before starting the pulsed sequence over again.

\begin{figure}[t!] 
	\centering
	\includegraphics[width=0.5\textwidth]{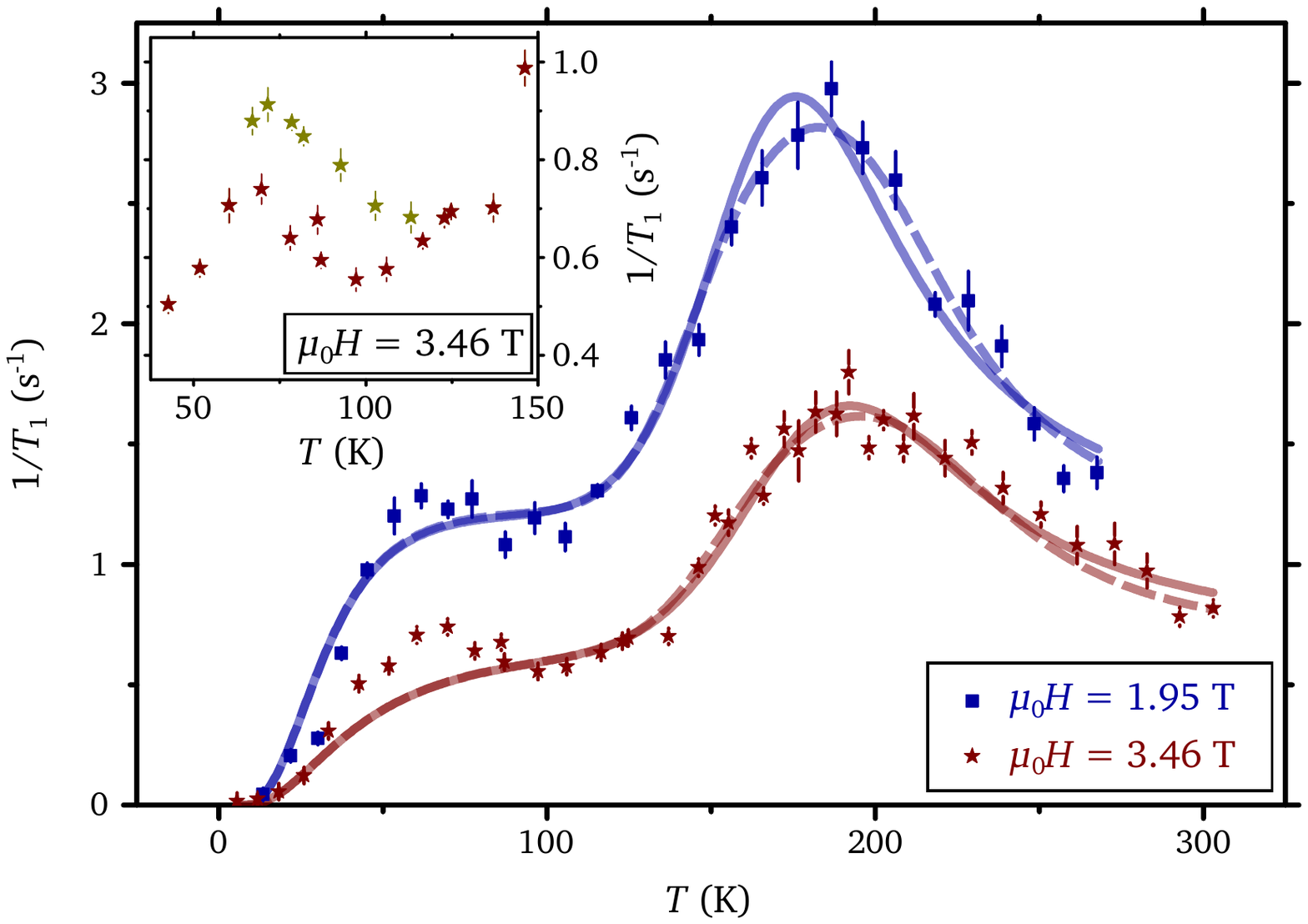}
	\caption{\label{1overT1vsT} Main panel: dependence of the spin-lattice relaxation rate on temperature at different values of the magnetic field. The continuous lines are the results of a global curve fitting based on Eq.~\eqref{EqBPPfitting}. The dashed lines are the results of a global curve fitting based on Eq.~\eqref{EqBPPfitting} assuming a flat distribution for the activation temperature $\vartheta_{\textrm{HT}}$ -- see Eq.~\eqref{EqDistrKT}. Inset: thermal-history effects on the spin-lattice relaxation rate observed for different cooling runs.}
\end{figure}

We report the dependence of the spectral width $w$ on temperature in Fig.~\ref{FWHMvsT}. The linewidth is approximately constant below $\sim 240$ K and it sharply gets much narrower at higher temperatures. In spite of the sizeable error bars, it is possible to detect a weak, field-independent increase of the linewidth on cooling the sample below $\sim 100$ K. The dependence of $1/T_{1}$ on temperature is shown in Fig.~\ref{1overT1vsT} for the two values of magnetic field. Although not shown explicitly, the stretching parameter takes values $\beta \simeq 0.9 \pm 0.05$ within the whole investigated temperature window without any detectable trend. As prominent features, we notice a local maximum at around $70$ K followed by a marked peak at around $180$ K. The amplitude as well as the position of these peaks are dependent on the magnetic field. In spite of the comparable thermal protocols used to perform our measurements, we stress that the spin-lattice relaxation rate shows effects of thermal history, as highlighted in the inset to Fig.~\ref{1overT1vsT}. In the following analysis, we will refer to the lower branch (red stars) because of an overall agreement shown between cooling cycles using liquid nitrogen and helium (with the exception of one specific cooling cycle using liquid nitrogen, denoted with yellow stars).

\section{Modelling the spin-lattice relaxation rate}\label{SectModelling}

The nuclear spin-lattice relaxation rate is proportional to the spectral density $J(\omega_{L})$ of the fluctuating local magnetic field at the nuclear site in the directions orthogonal to the quantization axis, and calculated at the Larmor angular frequency $\omega_{L} = 2\pi\nu_{L}$ \cite{Abr61,Sli90}. Assuming that the autocorrelation function for the fluctuations of the local field is an exponentially-decaying function leads to a conventional Lorentzian functional form for the spectral density \cite{Blo48}
\begin{equation}\label{EqSpectralDensity}
	J(\omega) = \frac{\tau_{c}}{1+\omega^{2}\tau_{c}^{2}}
\end{equation}
governed by the correlation time $\tau_{c}$ characterizing the dynamics. In the particular case of the relaxation being driven by the time-modulation of the nuclear dipole-dipole interaction, the following Kubo-Tomita (KT) formula is valid for the spin-lattice relaxation rate \cite{Kub54,Abr61}
\begin{equation}\label{EqKuboTomita}
	T_{1}^{-1}\left(T\right)^{\textrm{KT}} = C \left[J(\omega_{L}) + 4 \; J(2\omega_{L})\right]
\end{equation}
where the factor $C \sim \gamma^{2} \langle\Delta B^{2}\rangle$ includes the mean square amplitude of the transverse fluctuating field $\Delta B$ and the proton gyromagnetic ratio $\gamma$. Accordingly, we write
\begin{equation}\label{EqBPPfitting}
	T_{1}^{-1}\left(T\right) = T_{1}^{-1}\left(T\right)_{\textrm{LT}}^{\textrm{KT}} + T_{1}^{-1}\left(T\right)_{\textrm{HT}}^{\textrm{KT}}
\end{equation}
to account for the two independent processes observed at low and high temperature in the experimental data (LT and HT, respectively). Assuming $\tau_{c,\textrm{LT}} = \tau_{0,\textrm{LT}} \exp(\vartheta_{\textrm{LT}}/T)$ and $\tau_{c,\textrm{HT}} = \tau_{0,\textrm{HT}} \exp(\vartheta_{\textrm{HT}}/T)$, i.e., an Arrhenius-like dependence on temperature for both the correlation times with activation temperatures $\vartheta_{\textrm{LT}}$ and $\vartheta_{\textrm{HT}}$, gives us a fitting function to describe the experimental data in Fig.~\ref{1overT1vsT}.

The results of a global data fitting based on Eq.~\eqref{EqBPPfitting}, sharing all the fitting parameters except the Larmor frequencies between the two datasets, are shown in the main panel of Fig.~\ref{1overT1vsT} as continuous lines. The numerical results of the fitting parameters are reported in Tab.~\ref{TabFitting}. The agreement with the experimental data is satisfactory in spite of the deviation observed for $T \lesssim 80$ K for the high-field data. We stress that the slowing down of a time-modulated nuclear dipole-dipole interaction should be reflected in an extra-broadening of the spectral linewidth upon decreasing temperature based on the expression \cite{Abr61}
\begin{equation}\label{EqAbragam}
	w^{2} = \frac{2 w_{0}^{2}}{\pi} \arctan\left(\pi w \tau_{c}\right),
\end{equation}
where $w_{0}$ is the rigid-lattice linewidth. We solved Eq.~\eqref{EqAbragam} using the fitting parameters estimated from the global-fitting procedure described above and, in particular, the numerical values defining $\tau_{c,\textrm{HT}}$. The result is shown by the continuous line in the main panel of Fig.~\ref{FWHMvsT}. In spite of the sizeable error bars, the experimental data are consistent with the expected trend. A similar increase in the linewidth is expected at even lower temperatures based on Eq.~\eqref{EqAbragam} using $\tau_{c,\textrm{LT}}$ -- however, our experimental data at low temperatures are not dense enough to resolve the expected behaviour.

The quality of the global fitting in the main panel of Fig.~\ref{1overT1vsT} can be improved by assuming a statistical distribution of correlation times $\tau_{c,\textrm{HT}}$. For this aim, a straightforward way to proceed is to assume a well-defined value for $\tau_{0,\textrm{HT}}$ and a flat, square distribution of activation energies $\vartheta_{\textrm{HT}}$ extended over the window $\left(\vartheta_{\textrm{HT}}-\delta\vartheta_{\textrm{HT}},\vartheta_{\textrm{HT}}+\delta\vartheta_{\textrm{HT}}\right)$. The analytical result is derived in Appendix~\ref{AppKTDistr}. Plugging Eq.~\eqref{EqDistrKT} for the HT process in Eq.~\eqref{EqBPPfitting} and repeating the global data fitting results in the dashed lines in Fig.~\ref{1overT1vsT}. The numerical results of the fitting parameters are reported in Tab.~\ref{TabFitting}.

\begin{table*}[t!]
	\begin{tabular}{c || c | c | c || c | c | c | c}
		{} & $C_{\textrm{LT}}$ ($10^{8}$ s$^{-2}$) & $\tau_{0,\textrm{LT}}$ ($10^{-10}$ s) & $\vartheta_{\textrm{LT}}$ (K) & $C_{\textrm{HT}}$ ($10^{8}$ s$^{-2}$) & $\tau_{0,\textrm{HT}}$ ($10^{-13}$ s) & $\vartheta_{\textrm{HT}}$ (K) & $\delta\vartheta_{\textrm{HT}}$ (K)\\
		\hline
		\hline
		\multirow{2}*{Current sample} & $4.4 \pm 0.1$ & $6.1 \pm 0.2$ & $65 \pm 1$ & $6.5 \pm 0.1$ & $12 \pm 1$ & $1210 \pm 20$ & $0$ (fixed)\\
		\cline{2-8}
		& $4.4 \pm 0.1$ & $6.2 \pm 0.2$ & $65 \pm 1$ & $8.3 \pm 0.2$ & $1.2 \pm 0.1$ & $1650 \pm 10$ & $320 \pm 20$\\
		\hline
		\hline
		Sample from Ref.~\cite{Pia22} & $41.2 \pm 0.4$ & $0.59 \pm 0.01$ & $24.0 \pm 0.5$ & $15 \pm 1$ & $13 \pm 5$ & $1040 \pm 60$ & $350 \pm 30$\\
		\hline
	\end{tabular}
	\caption{\label{TabFitting} Results of the fitting procedures based on Eqs.~\eqref{EqBPPfitting} and \eqref{EqDistrKT} for the $1/\/T_{1}$ data shown in Figs.~\ref{1overT1vsT} and \ref{1overT1diffBatches}.}
\end{table*}

\section{Discussion}\label{SectDisc}

We interpret the sharp decrease of the spectral linewidth upon increasing temperature above $\sim 240$ K as a motional-narrowing effect associated with the diffusion of the intercalated hydrogen ions. More precisely, we argue that the condition $2\pi w \tau_{c}^{diff} \gg 1$ (i.e., $\tau_{c}^{diff} \gg 5 \; \mu$s) is satisfied for $T \lesssim 240$ K, where $\tau_{c}^{diff}$ is the correlation time characteristic of the hydrogen diffusion throughout the lattice. To reinforce our interpretation, we notice that our results are in perfect qualitative agreement with previous continuous-wave NMR measurements on hydrogen-intercalated 1$T$-TiS$_{2}$ \cite{Rit86}. This latter material is of particular importance for our aims since it is isostructural to 1$T$-TiSe$_{2}$ and, interestingly, it does not show any transition to a CDW phase. Based on these arguments, we safely conclude that the hydrogen ions form a rigid lattice integral with the 1$T$-TiSe$_{2}$ host crystal lattice for $T \lesssim 240$ K, i.e., where we observe all the features that we are going to discuss below.

\begin{figure}[t!] 
	\centering
	\includegraphics[width=0.5\textwidth]{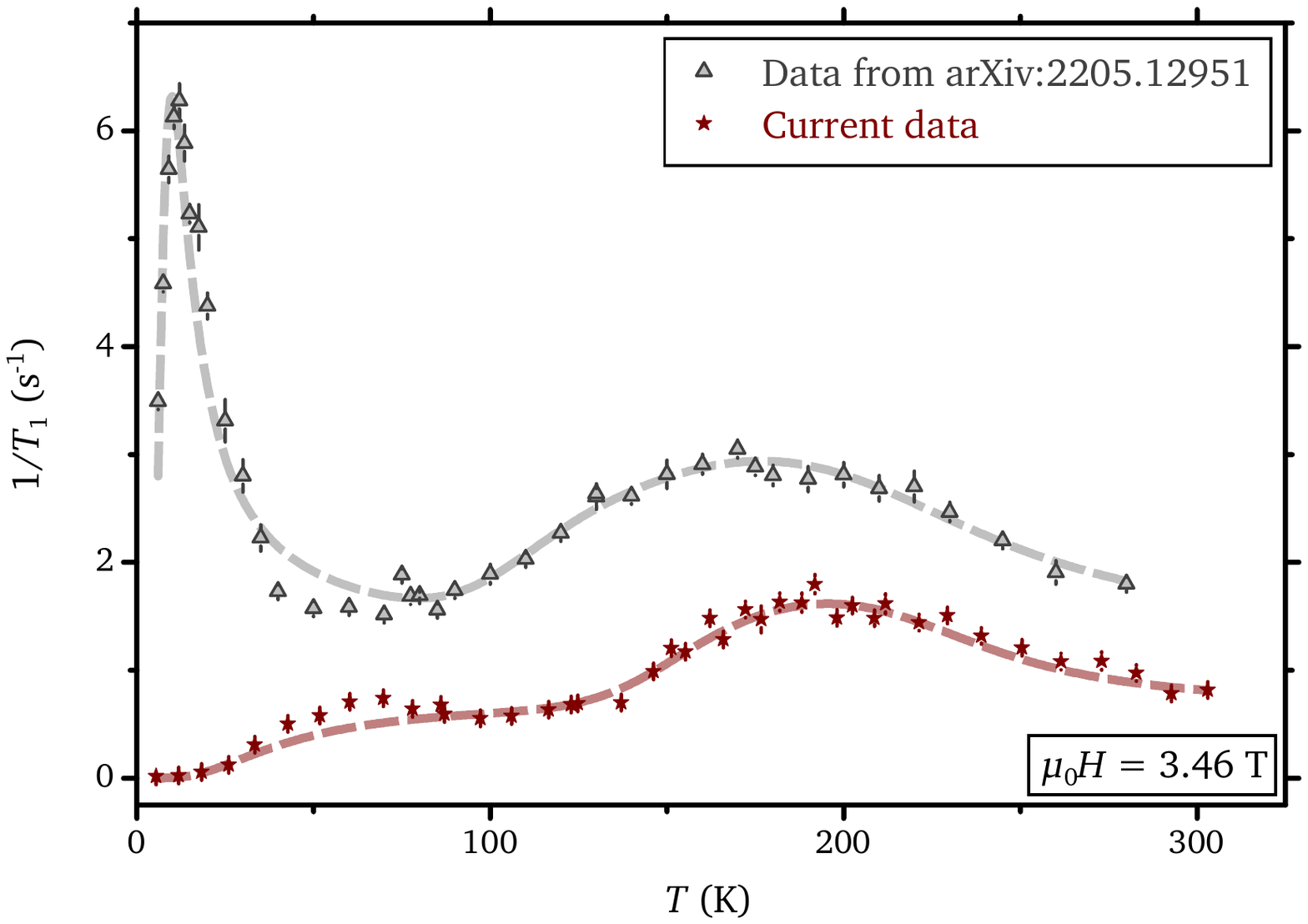}
	\caption{\label{1overT1diffBatches} Dependence of the spin-lattice relaxation rate on temperature at $\mu_{0}H = 3.46$ T for the current sample (red stars -- same data as in Fig.~\ref{1overT1vsT}) and for another hydrogen-intercalated TiSe$_{2}$ sample measured previously (grey triangles -- data from Ref.~\cite{Pia22}). The dashed lines are the best-fitting curves based on Eqs.~\eqref{EqBPPfitting} and \eqref{EqDistrKT}, assuming a statistical distribution of correlation times only for the activation temperature $\vartheta_{\textrm{HT}}$ (the red curve is the same as in Fig.~\ref{1overT1vsT}).}
\end{figure}
Before proceeding with the interpretation of our data, is interesting to compare the current results with the dependence of $1/T_{1}$ on temperature at $\mu_{0}H = 3.46$ T reported previously in Ref.~\cite{Pia22} for another collection of stacked protonated 1$T$-TiSe$_{2}$ crystals (gating time $30$ minutes -- see Fig.~\ref{1overT1diffBatches}). Our previous interpretation of those data was based on the Korringa-like linear trend between $80$ and $170$ K, suggesting that the deviations at higher temperatures could be due to a reconstruction of the Fermi surface resulting from the CDW state onset \cite{Pia22}. Such an interpretation was motivated by the observation of generalized Korringa trends both in pristine, isotopically-enriched TiSe$_{2}$ measured by {}$^{77}$Se NMR \cite{Dup77} as well as in other TMDs such as VSe$_{2}$, VS$_{2}$ and IrTe$_{2}$ \cite{Tsu81,Tsu83,Miz02}. The additional marked peak at lower temperature was reported for pristine TiSe$_{2}$ as well \cite{Dup77} -- although its origin has remained, to the best of our knowledge, elusive \cite{Nai92} -- clearly suggesting that our {}$^{1}$H NMR measurements are directly probing the intrinsic dynamics of TiSe$_{2}$. However, at this stage, the new currently discussed results urge us to reinterpret all the data within the same model. In particular, it is clear that the spin-lattice relaxation rate should be described for all the samples in terms of two well-defined maxima located at low and high temperatures, respectively, the latter seemingly correlating with $T_{\textrm{CDW}}$. The result of a best-fitting procedure based on Eqs.~\eqref{EqBPPfitting} and \eqref{EqDistrKT}, again assuming a flat distribution of correlation times for the HT process, is reported as a dashed line in Fig.~\ref{1overT1diffBatches} (and the resulting fitting parameters in Tab.~\ref{TabFitting}) showing a satisfactory agreement with the experimental data. In spite of the major quantitative differences between the two samples, we argue that the underlying mechanisms leading to the spin-lattice relaxation are the same.

\subsection{High-temperature peak}

We focus on the $T \gtrsim 100$ K limit first. In spite of the limited amount of experimental data, the results for pristine TiSe$_{2}$ in Ref.~\cite{Dup77} do not highlight any well-defined maximum in the spin-lattice relaxation rate for $T \gtrsim 100$ K, thus suggesting that the currently observed feature is specific to the hydrogen-intercalated system. The broad shape of the temperature dependence of $1/T_{1}$, together with the characteristic dependence on the magnetic field, suggest that the origin of the observed maximum should not be associated with any critical dynamics. Consistently with the observation on the narrowing of the spectral linewidth, it is unrealistic to associate such dynamics to the diffusion of the intercalated ions as, based on our comments above, the condition $1/\tau_{c}^{diff} \ll \omega_{L}$ is satisfied at least up to $\sim 240$ K. Consistently with these arguments, we stress that several systems with comparable structure and chemical formula show diffusion-related maxima in $1/T_{1}$ at higher temperatures if compared to our observations \cite{Skr89,Skr91a,Skr91b,Bow93,Kuc94,Wil06,Wil08}.

It is reasonable that the -- presumably sample-dependent -- intrinsic disorder associated with the intercalation process is reflected in a short-range ordered CDW phase \cite{Spe19}. Here, the order is confined within correlated mesoscopic islands lacking phase coherence among themselves at high temperatures until long-range order is recovered for $T \lesssim T_{\textrm{CDW}}$. The slow dynamics of the disordered state, corresponding to a CDW glass \cite{Nad95} or -- more precisely -- cluster glass, results in a non-critical peak in the spin-lattice relaxation rate consistent with our observations \cite{Jul99,Ura22}. The persistence of the contribution to $1/T_{1}$ at temperatures well above $T_{\textrm{CDW}}$ could imply that the intercalation process favours the insurgence of a segregated, preemptive CDW state at high temperatures, amplifying the tendency of the system towards the order implied by the partial gap opening for $T \gg T_{\textrm{CDW}}$ \cite{Miy95,Che16}.

Within this scenario, two different mechanisms can be identified in order to justify the relaxation mechanism for the nuclear ensemble. On the one hand, the dynamics of the CDW clusters would induce a slow, non-critical local dynamics of the lattice and of the intercalated hydrogen ions, in turn. In this view, a conventional nuclear dipole-dipole relaxation among the intercalated ions would be induced and drive the spin-lattice relaxation -- thus justifying the use of Eq.~\eqref{EqKuboTomita} above. Another possible mechanism is the time-modulation of the Knight shift distribution induced by the CDW state \cite{Bor77,Skr95}. The lack of any dependence of the extra linewidth broadening for $T \lesssim 150$ K on the magnetic field (see Fig.~\ref{FWHMvsT}), seems to rule out the latter mechanism in favour of the nuclear dipole-dipole relaxation.

\subsection{Low-temperature peak}

We now focus on the peak in $1/T_{1}$ observed for $T \lesssim 100$ K in Figs.~\ref{1overT1vsT} and \ref{1overT1diffBatches}. As mentioned above, this maximum as well as a frequency-dependent anomaly in the internal friction were observed in the same temperature range in pristine TiSe$_{2}$ \cite{Dup77,Bar77,Bar75}. This is a strong hint that the phenomenology observed by our experiments is not induced by the hydrogen intercalation but that it should rather be related to the intrinsic electronic properties of 1$T$-TiSe$_{2}$. Another interesting observation comes from copper-intercalated Cu$_{x}$TiSe$_{2}$. It is well known that increasing the amount of intercalated copper acts as an efficient knob to progressively weaken the CDW state until its full suppression for $x \simeq 0.06$ while superconductivity develops for $x \gtrsim 0.04$ \cite{Mor06}. This said, it is remarkable that the spin-lattice relaxation rate measured by {}$^{77}$Se NMR in two Cu$_{x}$TiSe$_{2}$ samples where the CDW state is strongly suppressed or even absent only show a well-defined Korringa-like linear dependence on temperature over the entire accessed range down to low temperatures \cite{Lum10}. Associating the well-defined low-temperature peak in $1/T_{1}$ to the intrinsic properties of the CDW is consistent with this latter observation. We stress that the local detection of intrinsic CDW features at the hydrogen site confirms that the phase segregation of CDW-ordered, hydrogen-poor regions and superconducting, hydrogen-rich regions -- with progressive volume redistribution within the two as the protonation time increases -- is not the correct model to describe the state realized by the intercalation, as already argued in \cite{Pia22}.

Having excluded that the low-temperature maximum in $1/T_{1}$ arises from a change in the electronic structure or from the onset of magnetism \cite{DiS76,Dup77,Woo76,Guy82}, a possible explanation for our results is that the spin-lattice relaxation is influenced by the low-frequency dynamics of the CDW itself. Indeed, the collective excitations of the CDW state (i.e., phase- and amplitude-modes) can influence the spin-lattice relaxation \cite{Ove71,Ove78,Bak82} -- however, some words of caution are in order. The vast majority of the results have been obtained using quadrupolar nuclei that couple directly with the local electric-field gradients and lattice distortions generated by the CDW state \cite{Bli81,Zum81,Bli83,Kog84,Pap91,Pap93,Mat99}. In our current case of {}$^{1}$H-NMR, we argue that the mechanism underlying the spin-lattice relaxation is due to a time-modulated nuclear dipole-dipole interaction due to a time-modulated lattice distortion induced by the CDW dynamics (see above).

At the same time, it is well-known that collective phase-mode excitations are not free to slip throughout the system in the case of a commensurate CDW state \cite{Bak82} akin to the one realized in TiSe$_{2}$, where the lattice is particularly efficient in pinning the spatial modulation of the charge. For this reason, we assume that the CDW state is localized but still able to oscillate around its equilibrium position as a result of thermal excitations. In this case, the time-dependent oscillations of the CDW can induce an efficient relaxation when the characteristic correlation time of the dynamics matches the inverse of the Larmor frequency. We expect that this phenomenology is strongly influenced by the level (and, more in general, the strength) of impurities, structural defects, and disorder in the system \cite{Hil14,Wu15,Hil18,Liu21,Fen23}, acting as pinning centres for the CDW state. Both the sample-dependent position and amplitude of the low-temperature peak as well as the observation of clear thermal-history hysteretic effects are consistent with this expectation \cite{Wan96}. Based on this interpretation, the observation of a higher $\vartheta_{\textrm{HT}}$ value for the current data -- if compared to the results reported in Ref.~\cite{Pia22} -- suggest that the currently-investigated sample contains a much higher degree of impurities and structural defects. This expectation is seemingly confirmed by the dependence of the electrical resistivity ($\rho$) on temperature (see Ref.~\cite{Pia22}) and by the residual resistivity ratio ($RRR$), $\rho(295 \; \textrm{K})/\rho_{0}$, characteristic of the two samples ($\rho_{0}$ being the residual resistivity just before superconductivity sets in). $RRR \simeq 9.3$ for the current sample while $RRR \simeq 12.1$ for the sample studied in Ref.~\cite{Pia22}, consistently with a lower degree of impurities and defects in the latter case.

\section{Summarizing remarks and conclusions}\label{SectConcl}

We discussed the results of {}$^{1}$H nuclear magnetic resonance experiments on hydrogen-intercalated TiSe$_{2}$ crystals, where the intercalation induces robust and non-volatile superconductivity \cite{Pia22}. Our results suggest that the disorder associated with the intercalation induces an inhomogeneous charge-density wave phase. In particular, we argue that stable mesoscopic charge-density-wave-ordered regions lacking phase coherence among them nucleate already at temperatures higher than the bulk transition temperature $T_{\textrm{CDW}}$. The non-critical dynamics of this cluster-glass charge-density wave state influences the spin-lattice relaxation of the {}$^{1}$H nuclear magnetization, showing a well-defined Kubo-Tomita-like trend with a characteristic dependence on temperature and on the magnetic field. Additionally, we reported clear evidences of an additional anomaly in the spin-lattice relaxation rate at low temperatures akin to what was reported for the pristine TiSe$_{2}$ composition in Ref.~\cite{Dup77} and whose origin should be related to the intrinsic properties of the charge-density wave phase. The low-temperature dynamics is strongly sample-dependent and is likely associated with the complicated interplay of the charge-density wave state with pinning centres such as impurities and structural defects. Further complementary insights into the charge-density wave dynamics could be obtained performing {}$^{7}$Li nuclear magnetic resonance experiments on lithiated TiSe$_{2}$, exploiting the non-vanishing quadrupole moment of {}$^{7}$Li nuclei and its direct coupling to local electric-field gradients and lattice distortions generated by the charge-density wave phase.

\section*{Acknowledgements}

We acknowledge insightful discussions with F. Borsa and A. Rigamonti. We thank S. Resmini for support during the NMR measurements.

\appendix

\section{Generalization of the Kubo-Tomita expression in the presence of a constant distribution of activation temperatures}\label{AppKTDistr}

The well-known Kubo-Tomita expression for the spin-lattice relaxation induced by the nuclear dipole-dipole interaction can be written as follows \cite{Kub54}
\begin{equation}\label{EqKTapp}
	\frac{1}{T_{1}} = C \; \left(\frac{\tau_{c}}{1+\omega_{L}^{2}\tau_{c}^{2}} + \frac{4\tau_{c}}{1+4\omega_{L}^{2}\tau_{c}^{2}}\right)
\end{equation}
with contributions from the spectral density $J(\omega)$ for the fluctuations of the local magnetic field perpendicular to the quantization axis calculated at $\omega_{L}$ and $2\omega_{L}$. Assuming an Arrhenius-like temperature dependence for both the correlation time $\tau_{c} = \tau_{0} \exp(\vartheta/T)$ and substituting in Eq.~\eqref{EqKTapp}, it is straightforward to show that
\begin{eqnarray}
	\frac{1}{T_{1}} &=&\frac{C}{2\omega_{L}} \left\{\left[\cosh\left(\frac{\vartheta}{T} + \ln(\omega_{L} \tau_{0})\right)\right]^{-1}\right. \nonumber\\ &&\left.+ 2 \; \left[\cosh\left(\frac{\vartheta}{T} + \ln(2\omega_{L} \tau_{0})\right)\right]^{-1}\right\}.
\end{eqnarray}

The effect of a distribution of correlation times (and, in particular, of activation energies) can be accounted for by considering the expression
\begin{eqnarray}
	\frac{1}{T_{1}} &=& \frac{C}{2\omega_{L}} \int_{-\infty}^{+\infty} \left\{\left[\cosh\left(\frac{\Delta}{T} + \ln(\omega_{L} \tau_{0})\right)\right]^{-1}\right. \\ &&\left.+ 2 \; \left[\cosh\left(\frac{\Delta}{T} + \ln(2\omega_{L} \tau_{0})\right)\right]^{-1}\right\} p(\Delta) d\Delta\nonumber
\end{eqnarray}
where $p(\Delta)$ is a normalized probability distribution function. An approximated but convenient choice is that of a constant distribution such that $p(\Delta) = 1/(2 \delta\vartheta)$ for $\vartheta-\delta\vartheta < \Delta < \vartheta+\delta\vartheta$ (and $p(\Delta) = 0$ elsewhere). Consequently,
\begin{eqnarray}
	\frac{1}{T_{1}} &=& \frac{C}{4\omega_{L}\delta\vartheta} \int_{\vartheta-\delta\vartheta}^{\vartheta+\delta\vartheta} \left\{\left[\cosh\left(\frac{\Delta}{T} + \ln(\omega_{L} \tau_{0})\right)\right]^{-1}\right. \nonumber\\ &&\left.+ 2 \; \left[\cosh\left(\frac{\Delta}{T} + \ln(2\omega_{L} \tau_{0})\right)\right]^{-1}\right\} d\Delta
\end{eqnarray}
which is solved as
\begin{eqnarray}\label{EqDistrKT}
	\frac{1}{T_{1}} &=& \frac{C \; T}{4 \omega_{L} \delta\vartheta} \left\{ \arctan\left[\sinh\left(\frac{\vartheta+\delta\vartheta}{T}+\ln(\omega_{L}\tau_{0})\right)\right]\right. \nonumber\\ & & - \arctan\left[\sinh\left(\frac{\vartheta-\delta\vartheta}{T}+\ln(\omega_{L}\tau_{0})\right)\right]\\ & & + 2 \arctan\left[\sinh\left(\frac{\vartheta+\delta\vartheta}{T}+\ln(2\omega_{L}\tau_{0})\right)\right] \nonumber\\ & & - 2 \left.\arctan\left[\sinh\left(\frac{\vartheta-\delta\vartheta}{T}+\ln(2\omega_{L}\tau_{0})\right)\right]\right\}. \nonumber
\end{eqnarray}

\providecommand{\noopsort}[1]{}\providecommand{\singleletter}[1]{#1}%

\end{document}